\documentstyle[epsfig]{mn2e}{} 
\begin{document}

\def\mpc{h^{-1} {\rm{Mpc}}} \def\up{h^{-3}
{\rm{Mpc^3}}} \def\uk{h {\rm{Mpc^{-1}}}}
\def\lsim{\mathrel{\hbox{\rlap{\hbox{\lower4pt\hbox{$\sim$}}}\hbox{$<$}}}}
\def\gsim{\mathrel{\hbox{\rlap{\hbox{\lower4pt\hbox{$\sim$}}}\hbox{$>$}}}}
\def\kms {\rm{km~s^{-1}}} \def\apj {ApJ} \def\aj {AJ} \def\mnras {MNRAS}
\def\aap {A\&A}

\title[Galaxies at void walls]{Low and High Surface Brightness Galaxies at Void Walls}

\author[L. Ceccarelli et al.  
]{\parbox[t]{\textwidth}{L. Ceccarelli$^{1,2}$, R. Herrera-Camus$^{3}$,  D. G. Lambas$^{1,2}$, G. Galaz$^{4}$ \& N. D. Padilla$^{4}$.
}
\vspace*{6pt}\\ $^1$ IATE, CONICET, Argentina.\\  
$^2$ Observatorio Astron\'omico de C\'ordoba, UNC, Argentina.\\  
$^3$ Department of Astronomy, University of Maryland, College Park, MD 20742, USA.
\\
$^4$ Departamento de Astronom\'\i a y Astrof\'\i ica,
Pontificia Universidad Cat\'olica de Chile, Santiago, Chile.
}

\date{\today}

\maketitle

\begin{abstract}
{
We study the relative fraction of low and high surface brightness
galaxies (LSBGs and HSBGs) at void walls in the SDSS DR7.  We focus on
galaxies in equal local density environments.  We assume that the host
dark-matter halo mass (for which we use SDSS group masses) is a good
indicator of local density.  This analysis allows to examine the
behavior of the abundance of LSBG and HSBG galaxies at a fixed local
density and distinguish the large-scale environment defined by the
void geometry. We compare galaxies in the field, and in the void
walls; the latter are defined as the volume of void shells of radius
equal to that of the void.  We find a significant decrement, a factor
$\sim 4$, of the relative fraction of blue, active star-forming LSBGs in
equal mass groups at the void walls and the field. This decrement is
consistent with an increase of the fraction of blue, active
star-forming HSBGs. By contrast, red LSBGs and HSBGs show negligible
changes. We argue that these results are consistent with a scenario
where LSBGs with blue colors and strong star formation activity at the
void walls are fueled by gas from the expanding void regions. This
process could lead to LSBG to HSBG transformations.
}
\end{abstract}

\begin{keywords} large scale structures: voids, galaxy groups,
statistical, LSB galaxies \end{keywords}

\section{Introduction}

\begin{figure*}
   \begin{picture}(430,250)
      \put(-40,0){\psfig{file=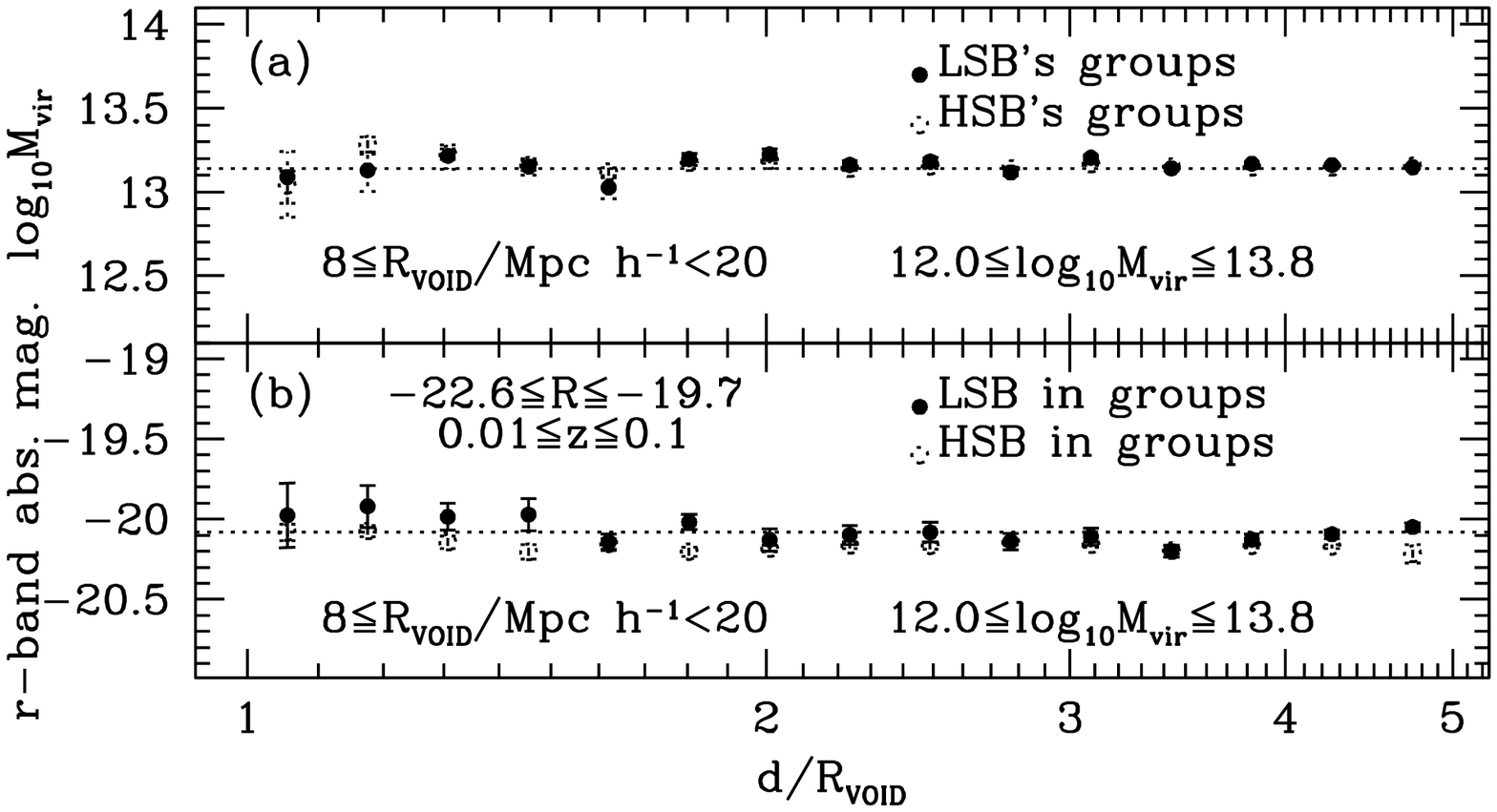,width=9.cm}}
      \put(220,0){\psfig{file=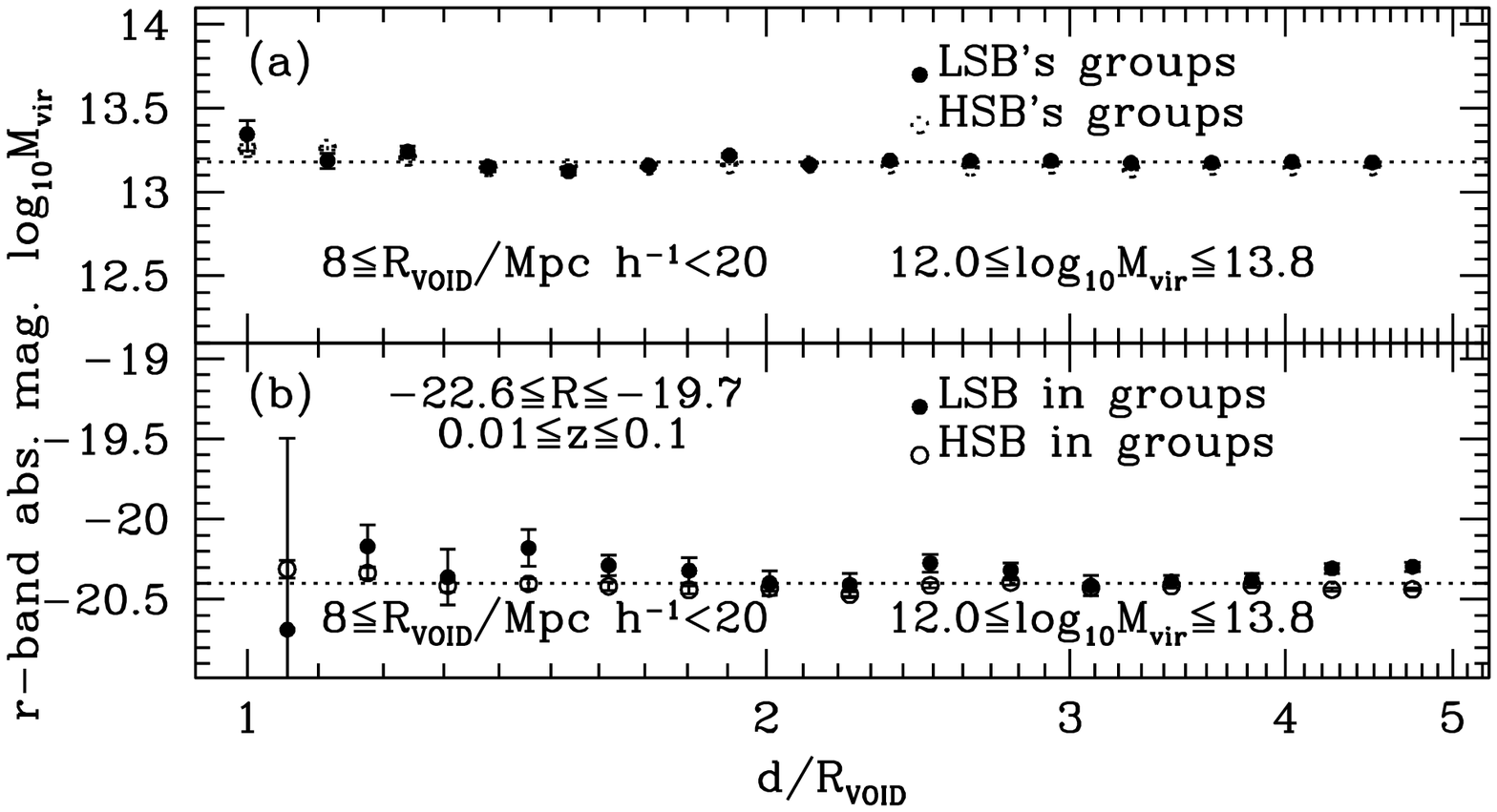,width=9.cm}}
   \end{picture}
   \vspace {-4.0cm}
   \caption
   { 
   {\it{Left}}
   (a) Median virial masses
   for selected groups hosting at least one blue, u-r $\leq 2.2$,
   LSB (filled circles) and HSB (open circles) galaxy as a function of
   normalized void-centric distance, for void radii in the range $8$
   h$^{-1}\leq $ R$_V < 20$ h$^{-1}$Mpc and group virial masses within 
   $10^{12.0}$ and $10^{13.8}$ M/M$_{\odot}$.
   (b) Median absolute magnitude of blue LSB (filled circles) and HSB
   (open circles) galaxies in groups as a function of normalized 
   void-centric distance, for void radii and group virial mases in 
   the same ranges. 
   {\it{Right}}
   (a) Median virial masses
   for selected groups hosting at least one star-forming, eclass $\leq -0.1$,
   LSB (filled circles) and HSB (open circles) galaxy as a function of
   normalized void-centric distance.
   (b) Median absolute magnitude of star-forming LSB (filled circles) and HSB
   (open circles) galaxies in groups as a function of normalized 
   void-centric distance.
   }
   \label{fig:f1}
\end{figure*}

The large-scale environment, and in particular void walls,
may have significant effects on the galaxy properties and their evolution.
It is well known that galaxy properties, such as morphology, star
formation rates and colours, vary strongly with the galaxy density in
their local environment (Dressler 1980; Lewis et al. 2002; Kauffmann et
al. 2004).  These variations are related to significant differences in how
the evolution of galaxies varies with the environment, 
mainly due to interactions,
merger histories, and assembly bias (e.g. Moore et al. 1998; Bell et
al. 2006).

It is expected that galaxies in voids have significantly different star
formation and chemical enrichment histories in comparison to those
of galaxies in denser environments (see, e.g., Peebles 2001; Gottl\"ober
et al. 2003; Hoeft et al.  2006; Hahn et al. 2007,
2009, and references therein).
It is well established that the highest surface brightness galaxies
reside in high density environments. However, 
the issue of the typical environment of low surface brightness galaxies
(LSBGs hereafter) is not completely settled  
albeit some authors suggest 
LSBGs populate preferentially lower density environments 
(see for instance Rosenbaum et al. 2009, Galaz et al. 2011)

LSBGs represent an important population among extragalactic objects. 
The central surface brightness of the disk in the B-band, $\mu_0$(B), 
is the photometric parameter typically used to separate the high and
the low surface brightness regime of galaxies. 
The most common threshold values found in the literature are between 
22 and 23 mag arcsec$^2$ (Impey et al. 2001).

LSBGs are characterized by many interesting observational properties
such as a low density of stars, which produce the low surface brightness. 
They also present extended flat rotation curves (de Blok  
2005; Swaters, Sanders \& McGaugh 2010)
having one of the highest M/L ratios in the Universe (Sprayberry et al. 1995).

The low star formation rate in combination with their rather isolated 
location in the cosmic web (Rosenbaum et al. 2009), as reported by several authors,
give clues for the understanding of their formation and evolution.
Several results provide evidence that 
large scale underdense regions, like cosmological voids, are characterized by  
coherent outflows of mass and galaxies moving
toward the void edges  (Ceccarelli et al., 2006, Padilla et al., 2005).
Consequently galaxies at void edges or walls
have undergone different evolutionary and merger histories than their field
counterparts. This could be due, for instance, to the void material 
accumulating around them, or to the fact that void galaxies most likely 
spent their lives inside voids.

Previous results have shown that galaxies at void walls present particular 
properties, such as in Ceccarelli, Padilla \& Lambas (2008).
Several studies have been focused on the properties of galaxies in
underdense regions. The luminosity function of galaxies in voids has
been measured by Rojas et al. (2005), who also study their photometric
properties finding that the population of galaxies in voids is
characterised by a fainter characteristic luminosity although the
relative importance of faint galaxies is similar to that found in the
field. Spectroscopic properties of void galaxies have also been studied
in detail (Hoyle, Vogeley \& Rojas, 2005); these results indicate that
galaxies inside voids have higher star formation rates than galaxies
in denser regions and are still forming stars at the same rate than in
the past.

Several authors suggest that LSBGs would be more isolated than HSBGs at
small scales (less than 2 Mpc, Bothun et al. 1993), 
and also between 2 Mpc and 5 Mpc (Rosenbaum et al. 2009). 
More recently, Galaz et al. (2011) found a deficit of neighbors for LSBGs 
at very small scales (less than 1 Mpc). 
These results motivate the
formulation of the following questions: Is the large-scale low density
environment, characteristic of voids, important for the set of LSBG 
properties? Can we find any signature in LSBGs residing in void walls?

{
Motivated by these facts, in this letter we perform a statistical
study of galaxies in void wall and in the field in the SDSS, 
ensuring that their local environments traced by the mass of 
their host groups is the same, and analysing the fraction of blue, 
star-forming galaxies in equal local environments given the
well known dependence on luminosity/stellar mass and local density
(Balogh et al., 2004, Baldry et al., 2006, Dekel \&
Birnboim, 2006, Kannappan 2004, Lagos et al., 2008).
}

\section{Data samples}

\subsection{LSB and HSB galaxies} 

{
 The galaxies studied in this work were extracted from the Main Galaxy Sample 
(Strauss et al. 2002) of the Sloan Digital Sky Survey data release 7 (SDSS DR7).
The SDSS is the largest survey carried out so far covering $10^4$~square~degrees 
and containing CCD imaging data in five photometric bands 
($ugriz$, Fukugita et al., 1996). The SDSS DR7 
spectroscopic catalogue (Abazajian et al. 2009) comprises 929,555 galaxies with 
a limiting magnitude of $r~\leq~17.77$~mag. Within this magnitude limit, 
Strauss et al. (2002) find that 99\% of the galaxies have half-light surface 
brightness brighter than 23~mag~arcsec$^{-2}$ in $r$ and only $\sim$0.01\% of 
galaxies are fainter than 24.5~mag~arcsec$^{-2}$ in $r$.

For a detailed description of how we select our sample we refer to Galaz et al. (2011). 
Essentially, we select late-type ($fracDevr\leq0.9$), nearly face on ($b/a<0.4$) 
galaxies in the redshift range  $0.01\leq~z~\leq0.1$ (we choose the lower cut to 
prevent peculiar velocity problems). For each galaxy, we calculate the surface 
brightness in the $B$-band using the band conversion from Smith et al. (2002). 
We correct the surface brightness for cosmological dimming. Finally, we use 
$\mu_{0}(B)~=~22.5$~mag~arcsec$^{-2}$ as the surface brightness cut to 
distinguish between LSB and HSBGs.}

Since the SDSS is a magnitude-limited survey
the observed populations of galaxies are not the
same at different redshifts. This prevents us to compare in a statistical 
way properties of nearby galaxies with those situated at further distances. 
In fact, the faintest galaxies are only registered at small redshifts while 
the brightest galaxies cover the whole catalogue. This magnitude-limited 
selection introduces a luminosity bias which can be removed by constructing 
a volume-limited catalogue, defining redshift and 
luminosity ranges where the absolute magnitude distributions are the same for 
all the galaxy populations.
A volume-limited catalogue allows us to compare two populations of LSBGs and 
HSBGs having the same absolute magnitude range, which for z $\leq$ 0.1 is 
M$_r \geq$ -19.7.

\subsection{Environmental information from SDSS galaxy group membership}

We use the galaxy group catalogue of Zapata et al. (2009) in order to
characterize the local environment of the galaxies in our samples.  As
was shown by Gonz\'alez \& Padilla (2009), Pasquali et al. (2009), and
Padilla, Lambas \& Gonz\'alez (2010), the mass of the host dark-matter halo
is one of the best tracers of variation of galaxy properties with the
environment.  Therefore, we choose this parameter to this end, and use
the luminosity of the four brightest members of the host group as a
tracer of its mass (see for instance, Eke et al., 2004).

The groups in the Zapata et al. catalogue are identified using a friends of friends
algorithm with varying projected linking length $\sigma$ with
$\sigma_0=0.239 $h$^{-1}$Mpc, and a fixed radial linking length
$\Delta v=450$ kms$^{-1}$, which provide a $95$ percent complete sample
with $<8$ percent contamination.  The minimum number of members per
group is $10$, with virial masses calculated using the gapper
estimator for the virial radius.

The dark-matter mass of the host halo of a galaxy is that of a galaxy
group that lies within a projected distance of
$2 $h$^{-1}$Mpc and a velocity difference of $800$ kms$^{-1}$.  
If no group satisfies this condition, the host halo mass
is assumed to fall below the completeness limit of the group catalogue.

\subsection{Identification of voids in the galaxy distribution}

We define a void as the largest spherical volume
within which the density is below a critical value
and we follow the procedure  
described in Ceccarelli et al. (2006)
to identify them in galaxy catalogues.
The algorithm used to found voids can be briefly 
described as follows: 
First, we
set a large number of random positions (for void centre candidates)
distributed throughout the catalogue. For each random position we consider 
the larger sphere 
satisfying the condition $\delta_{gx} < \delta_{max}$, where 
$\delta_{max}$ is the maximum density contrast allowed, and 
all the spheres considered are selected as void candidates. 
Finally, if there are superimposed
void candidates, all of them are removed and the only one considered
as a void is the largest underdense sphere, which should contain
the others. 

We apply the void finding algorithm to our volume limited sample of SDSS DR7 
galaxies.  
{
We set $\delta_{max}=-0.9$,
this value is in agreement with the mean density contrast 
of voids identified in the literature (Hoyle \& Vogeley 2004, 
Patiri et al. 2006a). However, small variations on $\delta_{max}$ 
do no affect the void statistics (Padilla et al. 2005).}
We have adopted $z=0.10$ as the limiting redshift of our sample. 
This choice of maximum redshift is a compromise between well-resolved voids
which require faint galaxies, and a sufficiently large volume in order to
have enough void statistics.  The adopted absolute magnitude limits imply
that the galaxy number density is high enough to lower the effects of
shot noise in the identification of small voids.  Our resulting sample,
containing $184$ voids, is restricted to radii within the range $8$
to $20$h$^{-1}$Mpc, which comprises the best resolved systems suitable
for our study.
We use the normalized void-centric distance (d/r$_{\rm void}$) and define 
the void walls by the range $0.9<$d/r$_{\rm void}<1.2$.

\section{LSB galaxies in groups at void walls}

\begin{figure*}
   \begin{picture}(330,380)
      \put(0,135){\psfig{file=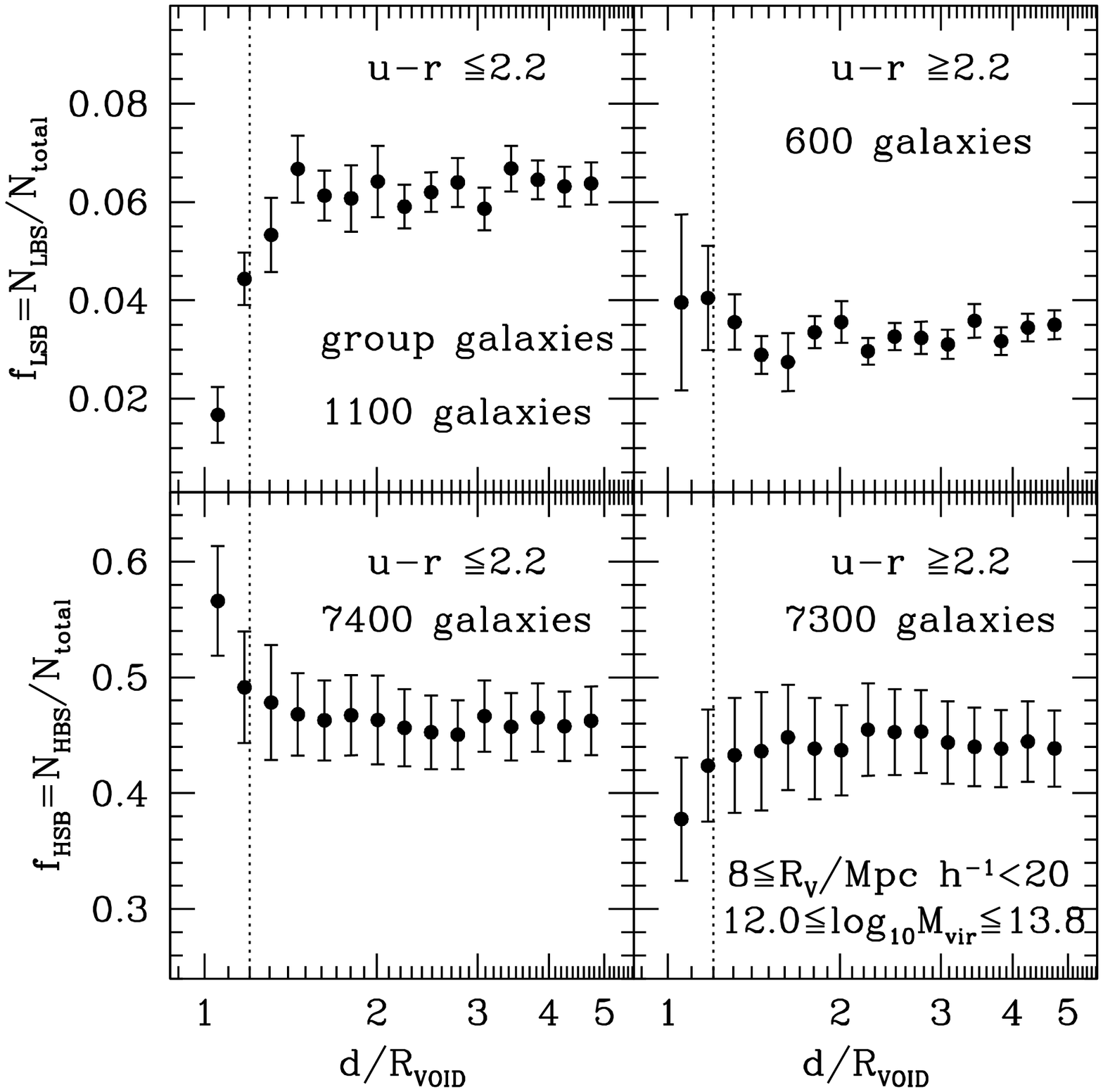,width=09.cm}}
      \put(0,-162){\psfig{file=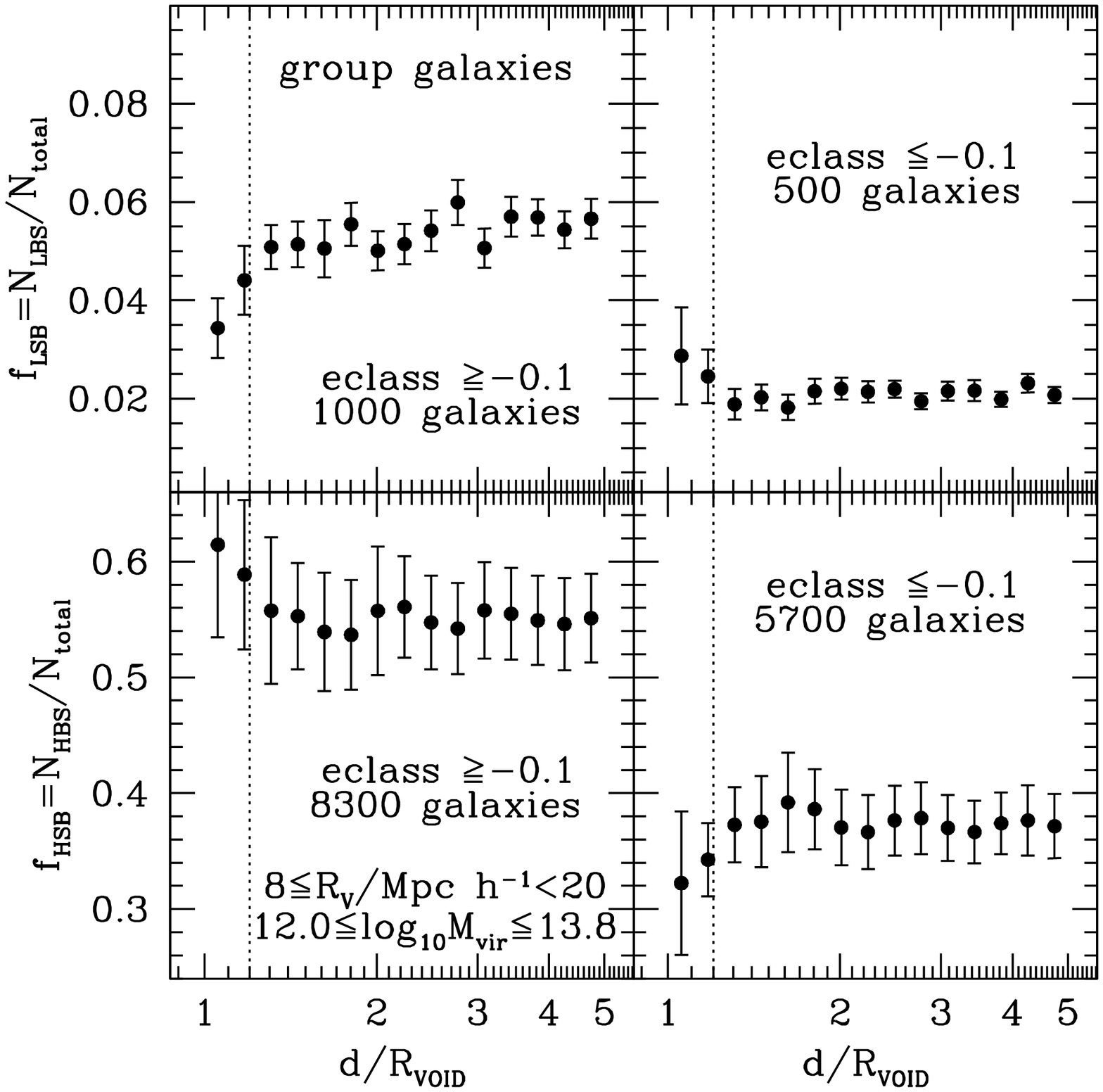,width=09.cm}}
   \end{picture}
   \vspace {-5.1cm}
   \caption
   { 
   Left: Relative fraction of blue,
   u-r $\leq 2.2$, LSB (upper panel) and HSB (lower panel)
   galaxies in groups as a function of normalized void-centric distance.
   Right: Relative fraction of red, u-r $\geq 2.2$, LSB (upper
   panel) and HSB (lower panel)
   galaxies in groups as a function of normalized void-centric distance.
   Void radii are in the range $8$ h$^{-1}\leq $ R$_V < 20$ h$^{-1}$Mpc
   and group virial masses within $10^{12.0}$ and $10^{13.8}$ M/M$_{\odot}$.
   The dotted lines indicate the outer boundary of the void wall. 
   }
   \label{fig:f2}
   \vspace {08.6cm}
   \caption
   {
   Left: Relative fraction of star-forming,
   eclass $\geq -0.1$, LSB (upper panel) and HSB (lower panel)
   galaxies in groups as a function of normalized void-centric distance.
   Right: Relative fraction of non star-forming, eclass $\leq -0.1$, LSB (upper
   panel) and HSB (lower panel)
   galaxies in groups as a function of normalized void-centric distance.
   Void radii are in the range $8$ h$^{-1}\leq $ R$_V < 20$ h$^{-1}$Mpc
   and group virial masses within $10^{12.0}$ and $10^{13.8}$ M/M$_{\odot}$.
   The dotted lines indicate the outer boundary of the void wall. 
   } 
\label{fig:f3}
\end{figure*}

\subsection{Properties of group LSB and HSB galaxies as a function of voidcentric distance}
{
Given the well documented dependence of galaxy properties on local
density, we study galaxies in equal mass groups, as this ensures an
equal local environment (Pasquali et al., 2009, Padilla, Lambas \&
Gonz\' alez, 2010).} 
We select group samples with equal distributions of virial masses 
at different void-centric distances. These samples allow an analysis of 
the properties of LSBGs in groups at void walls and a proper comparison 
to LSBGs in similar environments beyond the void walls. This procedure
should help to disentangle the relative weights of large scale (void
walls) and local effects (group) on LSBG properties.  
We use k-corrected $u-r$ colors (to $z=0.1$) and the parameter $eclass$ 
to classify the galaxies.  $eclass$
represents the projection of the first three principal 
components of the galaxy spectrum, and ranges from
-0.35 to 0.5 corresponding to the sequence of passive to strongly star forming  galaxies.
The top left panel in Figure \ref{fig:f1} shows the mean virial mass of 
SDSS DR7 selected groups containing at least $1$ blue 
($u-r\leq 2.2$)
\footnote{The color and $eclass$ adopted cuts 
divide the samples into the red and blue (and active and 
passive star formation) populations.} 
LSBG (HSBG) in filled (open) circles, as a function of normalized distance 
to the void centre. The mean virial mass of SDSS DR7 selected groups 
containing at least $1$ star-forming LSBG or HSBG ($eclass \leq -0.1$) 
is shown in the top right panel. As can be seen the mean virial mass 
of the samples shows similar values over the full range of void-centric 
distances. This assures that, on average, the local environments of our 
sample of LSBGs remain similar across the void walls.

Given that the distribution of LSB galaxies could exhibit a dependence on
luminosity and local density, we have explored the mean $M_r$ magnitude
of LSBGs for group galaxies in void walls and outside voids separately.
The bottom left panel in Figure \ref{fig:f1} shows the mean r-band 
absolute magnitude for blue ($u-r\leq 2.2$) LSBG (filled circles) in groups as a 
function of normalized distance to the void centre; open circles correspond 
to blue HSBGs in groups. The corresponding mean magnitude for star-forming 
LSBGs and HBSGs is shown in the bottom right panel. As can be seen 
the mean luminosity of group LSBG and HSBG samples is nearly constant 
for galaxies in the void walls and in the field. Thus, differences in 
the properties of LSBGs and HSBGs should mainly be related to the 
astrophysical effects associated to the special star formation history of 
galaxies which today reside within void walls and the field. 

\subsection{Relative abundances of galaxies according to color and spectral type}

We analyse the relative fraction of LSBGs and HSBGs in groups 
at different distances from void centres, taking into 
account their k-corrected $u-r$ colors (to $z=0.1$).

The results are shown in Figure \ref{fig:f2}, where it can be appreciated 
that the sample of HSBGs (lower panels) shows the well documented behaviour
that bluer galaxies occupy preferentially the void neighbourhoods.

The left panels in Figure \ref{fig:f2} show the relative fraction of LSBGs
(upper panel) and HSBGs (lower panel) in groups, corresponding to blue
objects ($u-r < 2.2$). 
It can be seen that at the void walls  there is a strong systematic
drop in LSBG fraction (upper panel), whereas the fraction of HSBGs 
increases (lower left panel of Figure \ref{fig:f2}), at a fixed local 
density (Ceccarelli, Padilla \& Lambas 2008, Gonz\'alez \& Padilla 2009).  

The right panels in Figure \ref{fig:f2} show the relative fraction of LSBGs
(upper panel) and HSBGs (lower panel) in groups, corresponding to red
objects ($u-r > 2.2$). As  can be appreciated we obtain similar fractions 
of red LSB and HSB galaxies over the full range of void-centric distances.
From a more detailed analysis of the lower right panel of 
Figure \ref{fig:f2} it can be noticed a slight drop 
on the faction of red HSBGs in groups at void walls. However, in general, 
the mean fraction of group LSBG and HSBG samples is nearly constant for 
red galaxies in the void walls and in the field.

We have also examined the star 
formation activity of these LSB galaxies at void walls through the $eclass$ parameter. 
In Figure \ref{fig:f3} we present 
the relative fraction of LSB and HSB galaxies considering high and low 
$eclass$ parameter values separately. Again we have considered a division into passive and
star forming galaxies with the $eclass$ threshold  $ = -0.1$ .
In the left upper panel of Figure \ref{fig:f3} 
we show the fraction of star-forming galaxies for LSBGs 
in groups. As can be seen, we find a decrement in the fraction of 
star-forming galaxies at walls in comparison to the field. As it is 
expected, the fraction of star-forming HSB galaxies increases at the void 
walls (left lower panel of Figure \ref{fig:f3}), in agreement with 
the results from Figure \ref{fig:f2}.
This is consistent with the relative changes in the fractions of blue 
and red galaxies of Figure \ref{fig:f2}.

{
Given our careful selection procedure and tests of uniformity of the
local environment of the galaxies in our samples, we conclude that the
behaviour of low fraction of blue, star-forming LSBGs is an intrinsic
property of void walls. We argue that this can be associated to the
transformation of void-wall LSBGs into star-forming HSBGs possibly due
to the gas arriving from the void interior as a consequence of void
expansion (we refer readers to Padilla et al. 2005, and Ceccarelli et al. 2006, 
for evidence suggesting that material from voids reachs walls).
We will explore this and others scenarios in numerical
simulations in a forthcoming paper.}

\section{CONCLUSIONS}

{
We have performed a statistical study of the population of galaxies
at void walls in comparison to the field.  We studied their colors and
spectral classes.  Since these properties have been shown to depend on
environment and galaxy luminosity, we made sure that the immediately
local environment of void wall and field galaxies is the same, via a
selection of host group masses, and luminosities so that their average
values are matched.
This selection allowed us to focus only on the effects of the global
environment associated to the void walls. Our findings can be
summarized as follows:
i) There is a remarkable systematic decrease of blue, active
star-forming LSBG fractions in groups at void walls, even when their
luminosities and environment are the same.
ii) Under the same conditions, we find an increase of the fraction of
blue, active
star-forming HSBGs at void walls.
iii) There is also a mild decrease of the fraction of red, pas-sive 
star-forming HSBGs at void walls.
iv) We find no significant trend in the fraction of red,
passive star-forming LSBGs at void walls, for galaxies with equal
environment and luminosity.}

\section*{Acknowledgments} This work has been partially supported by
Consejo de Investigaciones Cient\'{\i}ficas y T\'ecnicas de la Rep\'ublica
Argentina (CONICET), the Secretar\'{\i}a de Ciencia y T\'ecnica de
la Universidad Nacional de C\'ordoba (SeCyT), 
GG is supported by Fondecyt regular 1120195 and NP by Fondecyt regular 1110328.


\begin{thebibliography}{}

\bibitem[Abazajian et al.,2009]{Abaz} 
Abazajian K.N. et al., 2009, ApJS, 182, 543.

\bibitem[Baldry et al.,2006]{baldry} Baldry, I. K.; Balogh, M. L.;
Bower, R. G.; Glazebrook, K.; Nichol, R. C.; Bamford, S. P.; Budavari,
T., 2006, MNRAS, 373, 469.

\bibitem[Balogh(2004)]{balogh04} Balogh, Michael L.; Baldry, Ivan K.;
Nichol, Robert; Miller, Chris; Bower, Richard; Glazebrook, Karl, 2004,
ApJ, 615, 101.

\bibitem[bell]{bell} Bell, E. et. al 2006, ApJ, 652, 270

\bibitem[Ceccarelli2006]{cecc06} Ceccarelli L., Padilla N.D.,
Valotto C., Lambas D.G., 2006, MNRAS, 373, 1440.

\bibitem[Ceccarelli2008]{cecc08} Ceccarelli L., Padilla N.D.,
Lambas D.G., 2008, MNRAS Letters, 390, 9.

\bibitem[Dekel]{Dekel} Dekel, A., \& Birnboim, Y., 2006, MNRAS, 368, 2.

\bibitem[Dressler]{Dressler} Dressler A. 1980, ApJ, 236, 351.

\bibitem[eke04]{eke04} Eke V. R. (The 2dFGRS Team) et al., 2004, MNRAS, 355, 769

\bibitem[Fuku]{Fuku} 
Fukugita, M., Ichikawa, T., Gunn, J.E., Doi, M., Shimasaku, K.,
\& Schneider, D. 1996, AJ, 111, 1748.

\bibitem[Galaz]{Galaz} 
Galaz G. Herrera-Camus R., Garcia-Lambas D., Padilla N.
2011, ApJ, 728, 74.

\bibitem[Gottlober (2003)]{gottl03} Gottl\"ober D. M., Lokas E. L., Klypin
A. \& Hoffman Y., 2003, \mnras, 344, 715.
 
\bibitem[Hahn07]{Hahn07} Hahn, O., Porciani C., Dekel A., Carollo C. M., 2007, MNRAS, 381, 41.
 
\bibitem[Hahn09]{Hahn09} Hahn, O., Carollo C. M., Porciani C., Dekel A., 2007, MNRAS, 398, 1742.

\bibitem[Hoeft (2006)]{Hoeft06} Hoeft M., Yepes G., Gottl\"ober S., Springel V.,
2006, MNRAS, 371, 401.

\bibitem[Hoyle (2005)]{Hoyle05} Hoyle F., Vogeley M. S., Rojas R. 2005,
ASS, 206, 1002.

\bibitem[Impey]{Impey} Impey, C, Burkholder, V, Sprayberry D. 2001, AJ, 122, 2341

\bibitem[Kannappan]{Kannappan} Kannappan, S., 2004, ApJ, 611, L89.
 
\bibitem[kauffmann]{Kauffmann} Kauffmann G. et. al, 2004, MNRAS, 353, 713

\bibitem[Lagos et al (2008)]{Lagos08} Lagos, C., Cora, S., \& Padilla,
N., 2008, MNRAS, submitted.
 
\bibitem[lewis]{lewis} Lewis I. et. al, 2002, MNRAS, 334, 673

\bibitem[moore]{moore} Moore B., Lake G., Katz N. 1998, ApJ, 495, 139

\bibitem[Padilla (2005)]{padil05} Padilla, N. D.; Ceccarelli, L.; Lambas,
D. G. 2005, \mnras, 363, 977.

\bibitem[Pasquali(2009)]{pasqu09}Pasquali A., van den Bosch F. C., Mo H. J., Yang X., Somerville R.,
2009, MNRAS, 394, 38.

\bibitem[Peebles (2001)]{peebles01} Peebles P. J. E. 2001, \apj, 557, 495.

\bibitem[Rojas (2005)]{rojas05} Rojas R., Vogeley M. S., Hoyle, F.,
Brinkmann J. 2005, \apj, 624, 571.

\bibitem[Rosenbaum]{Rosenbaum} Rosenbaum S. D., Krusch E., Bomans D. J., Dettmar R. J.
2009, A\&A, 504, 807.

\bibitem[Sprayberry]{Sprayberry} Sprayberry D., Bernstein G. M., Impey C. D., Bothun G. 
1995, ApJ, 438, 72.

\bibitem[Strauss]{Strauss} Strauss M. A. et. al. 2002, AJ, 124, 1810.

\bibitem[Swaters]{Swaters} Swaters R. A., Sanders R. H., McGaugh S. S. 2010, ApJ, 718, 380.

\bibitem [zap4]{Z04} Zapata T., Perez J., Padilla N., Tissera P., 2009, MNRAS, 394, 2229
\end{thebibliography}
\end{document}